\documentclass[twocolumn,floatfix]{aastex631}

\newcommand{\ra}[1]{\renewcommand{\arraystretch}{#1}}

\usepackage{amsfonts}
\usepackage{booktabs}
\usepackage{savesym}
\savesymbol{tablenum}
\savesymbol{splitbox}
\usepackage{siunitx}
\restoresymbol{SIX}{tablenum}

\usepackage{float}

\usepackage{adjustbox}
\restoresymbol{SIX}{splitbox}
\usepackage{rotating}
\usepackage{gensymb}
\usepackage{array,multirow,graphicx}

\shorttitle{MACS\,J1149 Magnification Measurements}
\shortauthors{Williams et al.}

\graphicspath{{./}{ }}

\begin{document}

\title{Sp1149 II: Spectroscopy of \ion{H}{2} Regions Near the  Critical Curve of MACS\,J1149 and Cluster Lens Models}

\date{September 2023}

\author[0000-0002-1681-0767]{Hayley Williams}
\affiliation{School of Physics and Astronomy, University of Minnesota 
116 Church Street SE, Minneapolis, MN 55455 USA 
}
\author[0000-0003-3142-997X]{Patrick Kelly}
\affiliation{School of Physics and Astronomy, University of Minnesota 
116 Church Street SE, Minneapolis, MN 55455 USA 
}

\author[0000-0003-1060-0723]{Wenlei Chen}
\affiliation{School of Physics and Astronomy, University of Minnesota 
116 Church Street SE, Minneapolis, MN 55455 USA 
}

\author[0000-0001-9065-3926]{Jose Maria Diego}
\affiliation{IFCA, Instituto de F\'{i}ısica de Cantabria (UC-CSIC), Av. de Los Castros s/n, 39005 Santander, Spain}

\author[0000-0003-3484-399X]{Masamune Oguri}
\affiliation{Center for Frontier Science, Chiba University, 1-33 Yayoi-cho, Inage-ku, Chiba 263-8522, Japan}
\affiliation{Department of Physics, Graduate School of Science, Chiba University, 1-33 Yayoi-Cho, Inage-Ku, Chiba 263-8522, Japan}

\author[0000-0003-3460-0103]{Alexei V. Filippenko}
\affiliation{Department of Astronomy, University of California, Berkeley, CA 94720-3411, USA}

\begin{abstract}
Galaxy-cluster gravitational lenses enable the study of faint galaxies even at large lookback times, and, recently, time-delay constraints on the Hubble constant. 
There have been few tests, however, of lens model predictions adjacent to the critical curve ($\lesssim$ 8$''$) where the magnification is greatest. 
In a companion paper, we use the GLAFIC lens model to constrain the Balmer $L-\sigma$ relation for \ion{H}{2} regions in a galaxy at redshift $z=1.49$ strongly lensed by the MACS\,J1149 galaxy cluster. Here we perform a detailed comparison between the predictions of ten cluster lens models which employ multiple modeling assumptions with our measurements of 11 magnified giant \ion{H}{2} regions. We find that that the models predict magnifications an average factor of $6.2$ smaller, a $\sim2\sigma$ tension, than that inferred from the \ion{H}{2} regions under the assumption that they follow the low-redshift $L-\sigma$ relation. 
%Here we perform a novel test of cluster lens models by directly measure the magnification due to the Hubble Frontier Fields cluster MACS\,J1149.5+2223 adjacent to its critical curve $(\sim 2 - 8'')$ using the Balmer $L-\sigma$ relation for giant \ion{H}{2} regions.
%At these small offsets from the critical curve, lens models predict magnifications up to $\sim 20$ and their uncertainties are greatest. 
%The same MACS\,J1149 models we test were recently used to constrain the value of $H_0$ from the multiple appearances of SN Refsdal. Using measurements of the observed H$\alpha$ luminosities and intrinsic velocity dispersions of 11 giant \ion{H}{2} regions in the multiply-imaged spiral galaxy Sp1149 at redshift $z=1.49$, we employ the Balmer $L-\sigma$ relation we measure for low redshift ($z \approx 0.03$) \ion{H}{2} regions to estimate the intrinsic H$\alpha$ luminosity of the \ion{H}{2} regions in Sp1149 and infer the magnification due to lensing at their positions. Sp1149 is the host galaxy of the first-known multiply-imaged supernova, SN Refsdal, which has been used to infer the value of $H_0$ from time delays. 
%We compare our measurements of 11 giant \ion{H}{2} regions with the predicted magnifications from 10 available cluster models and find 
To evaluate the possibility that the lens model magnifications are strongly biased, we next consider the flux ratios among knots in three images of Sp1149, and find that these are consistent with model predictions. 
Moreover, while the mass-sheet degeneracy could in principle account for a factor of $\sim 6$ discrepancy in magnification, the value of H$_0$ inferred from SN Refsdal's time delay would become implausibly small. 
We conclude that the lens models are not likely to be highly biased, and that instead the \ion{H}{2} regions in Sp1149 are substantially more luminous than the low-redshift Balmer $L-\sigma$ relation predicts. 
%I added the sentence above about Refsdal -- since we think that it's likely not the lens model probably doesn't make sense to say this, I think? If this result represents a true bias in the magnification maps of the MACS\,J1149 cluster, rather than a redshift evolution in the intercept of the $L-\sigma$ relation, the lens models used to infer the value of $H_0$ from SN Refsdal may require revision.   
\end{abstract}

\section{Introduction}

\begin{figure*}[ht]
\centering
\includegraphics[width=1\linewidth]{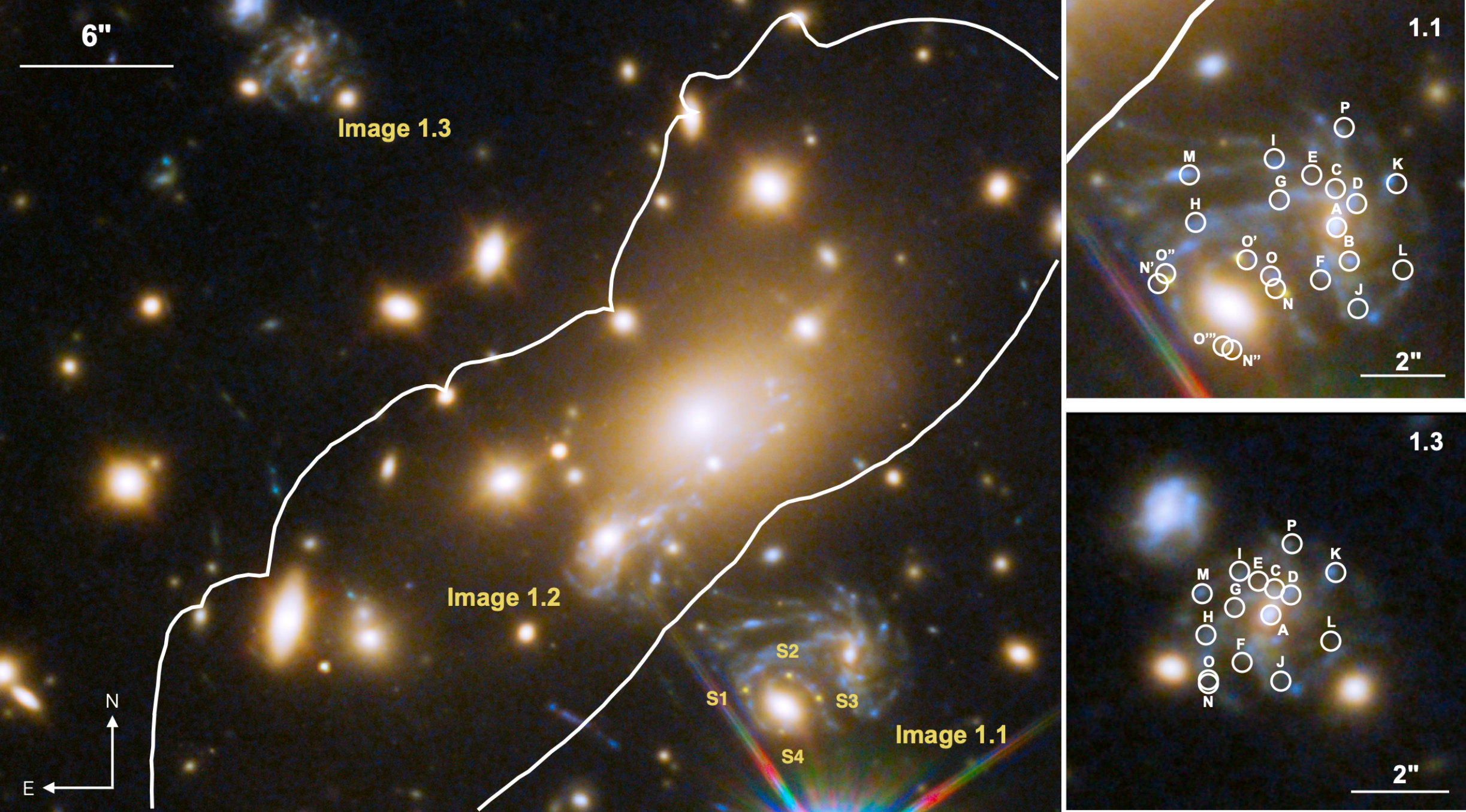}
\caption{False-color image of a portion of the MACS\,J1149 lensing cluster from {\it HST} imaging of the field. The three images of the face-on spiral galaxy Sp1149 are labeled, and the critical curve of the cluster (GLAFIC model) is shown as a white line. The appearances of SN Refsdal in an Einstein cross configuration are labeled S1--S4. Images 1.1 and 1.3 are shown in the right panels, with the \ion{H}{2} regions identified in white.}
\label{fig:macsj1149}
\end{figure*}

\begin{figure}[ht]
    \centering
    \includegraphics[width=\linewidth]{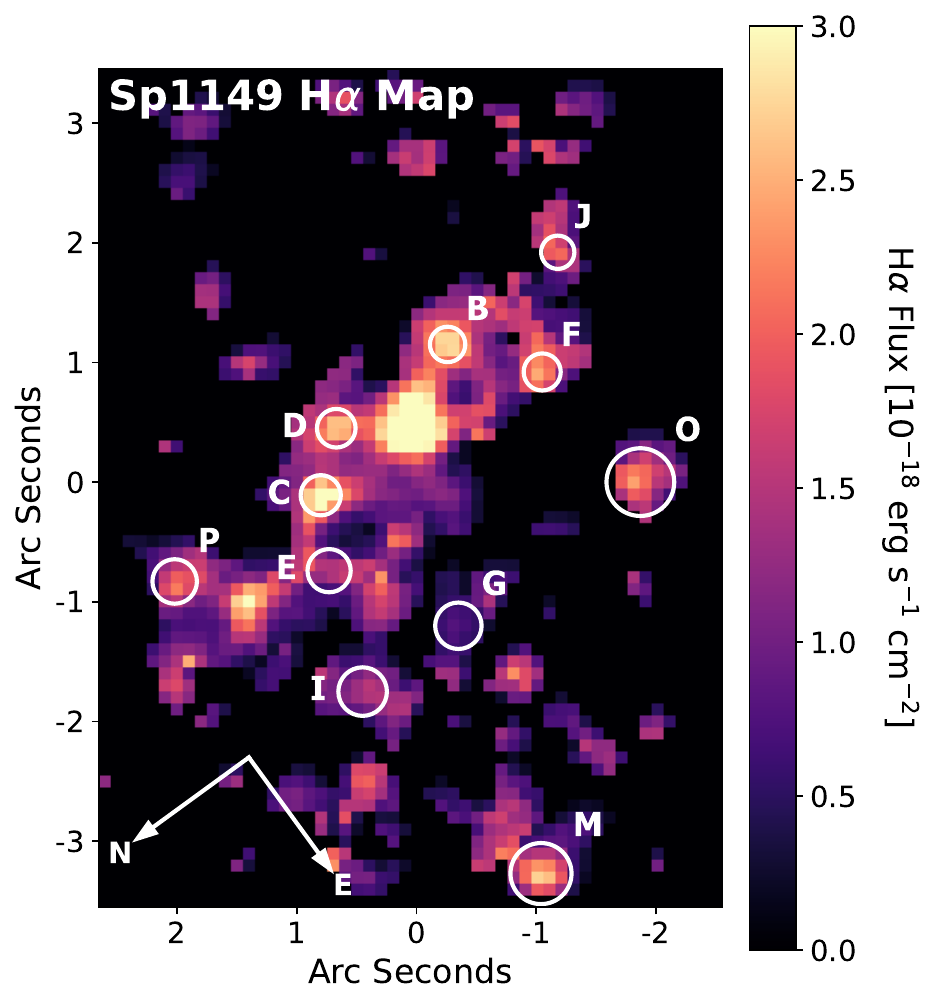}
    \caption{Emission-line intensity map of H$\alpha$ in Image 1.1 of Sp1149, with the 11 \ion{H}{2} regions we use to infer the magnification labeled in white. Labels are the same as Figure~\ref{fig:macsj1149}. This map was created from archival OSIRIS IFU observations of Sp1149 (PI: Kewley).}
    \label{fig:osiris_HAmap}
\end{figure}

Gravitational lensing by galaxy clusters provides a uniquely powerful tool for both finding and studying intrinsically faint galaxies and stars that existed when the Universe was a fraction of its present age. The magnifying power of galaxy-cluster lenses amplifies the flux and increases the angular sizes of background galaxies, enabling measurements that would otherwise not be possible \citep[e.g.,][]{Swinbank_2009,Wuyts_2014,Wang_2017,Curti_2020,Williams_2023}. \citet{Refsdal_1964} showed that, in principle, the time delays between multiple images of the same strongly-lensed supernova (SN) could be used to measure the Hubble constant (H$_0$). % \cite{Refsdal_1964}. 
This method of measuring H$_0$ using time delays provides an independent means to address the  ``Hubble tension'' between measurements of H$_0$ from Type Ia supernovae (SNe~Ia) in the local Universe \citep{Riess_2022} and from early-Universe observations of the cosmic microwave background \citep[CMB;][]{Planck_2020}. %with a $5\sigma$ statistical significance. 
If the Hubble tension represents a true difference, reconciling the inferred values may require revision to the standard $\Lambda$-cold-dark-matter ($\Lambda$CDM) cosmological model.

Uncertainties associated with cluster lens models limit our ability to study magnified populations of galaxies, as well as measure H$_0$ from time delays. Models use systems of multiple images of lensed background sources as constraints, yet also require assumptions about the connection between the distribution of luminous matter and dark matter. For well-constrained galaxy clusters such as the Hubble Frontier Fields \citep[HFF;][]{Lotz_2017}, the latest modeling techniques predict magnifications that are consistent among different models up to a magnification factor of $\mu \approx 40$--50 \citep{Bouwens_2022}. These modeling techniques have been tested on simulated images of galaxy clusters, and most models can reliably predict the magnification due to lensing within an accuracy of 30\% for magnifications up to $\mu \approx 10$ \citep{Meneghetti_2017}. Uncertainties become more pronounced in regions near the critical curve where $\mu \gtrsim 60$, where different lens modeling techniques can predict magnifications ranging from $\sim40$ to $\sim100$ at the same positions \citep{Bouwens_2022}. Identifying the most accurate lens modeling assumptions would improve our ability to use gravitational lenses as tools to study the galaxies that existed in earlier epochs of the Universe as well as measure H$_0$ from time-delay cosmography.  

%yield discrepancies up to $\sim 70\%$ in the predicted magnification \citep[e.g.,][]{Livermore_2017,Priewe_2017}. These uncertainties can make it difficult to properly interpret the extremely faint galaxies at high redshift that are only detectable when highly magnified. For example, \cite{Bouwens_2017} found that the large systematic uncertainties in magnification maps at $\mu\gtrsim30$ make it impossible to constrain the shape of the luminosity function of $z\sim6$ galaxies with $M_{\rm UV, AB} > -14$ mag. Identifying the most accurate lens modeling assumptions would improve our ability to use gravitational lenses as tools to study the galaxies that existed in earlier epochs of the Universe and derive $H_0$ from time delays. 

SN Refsdal, the first known multiply-imaged SN, was discovered in 2014 \citep{Kelly_2015,Treu_2016} in an Einstein-cross configuration around an elliptical galaxy in the Hubble Frontier Fields \citep[HFF;][]{Lotz_2017} galaxy cluster MACS\,J1149.5+2223 (redshift $z=0.544$, hereinafter referred to as MACS\,J1149), and reappeared $\sim 8''$ away in 2015 \citep{Kelly_2016}. SN Refsdal provided the first opportunity to use a cluster-scale gravitational lens to infer the value of H$_0$. The systematic uncertainties associated with cluster-scale lens models differ from the galaxy-scale models that have been used in previous measurements of H$_0$ from multiply-imaged quasars \citep{Grillo_2020}. The time delay between the 2014 and 2015 appearances of SN Refsdal was measured within 1.5\%  \citep{Kelly_2023}. Given a perfect lens model, this would yield an equally precise constraint on H$_0$. In the case of SN Refsdal, the greatest contribution to the uncertainty in H$_0$ is the uncertainty associated with the MACS\,J1149 cluster mass models \citep{Kelly_2023,Grillo_2020}. The value of H$_0$ derived from the time delays of SN Refsdal is H$_0 = 64.8^{+4.4}_{-4.3}$~km~s$^{-1}$~Mpc$^{-1}$ using the full set of pre-reappearance lens models, and $66.6^{+4.1}_{-3.3}$~km~s$^{-1}$~Mpc$^{-1}$\ using the two models that best reproduce the $H_0$-independent observables \citep{Kelly_2023}. %Intriguingly, this value is closer to the value of $H_0$ inferred from observations of the CMB than that measured from Type Ia SNe in the local Universe.

After the discovery of SN Refsdal, an individual, highly magnified ($\mu \approx 600$) blue supergiant star was discovered in the same host galaxy as SN Refsdal in the MACS\,J1149 field \citep{Kelly_2018}. Known as ``Icarus,'' the star was the first individual star discovered at a cosmological distance. An image of Icarus is always detectable in sufficiently deep images. The light curve of Icarus can place constraints on the abundance of primordial black holes \citep{oguridiegokaiser18} and the initial-mass function (IMF) of stars responsible for intracluster light, but inferences depend on the ability of lens models of the MACS\,J1149 cluster to predict the magnification of Icarus.

The MACS\,J1149 cluster lens has been modeled using several different methods. So-called simply parameterized models have been constructed by \cite{Sharon_2015},  \cite{Keeton_2010},  \cite{Grillo_2016}, the GLAFIC team \citep{Oguri_2010,Kawamata_2016}, and the Clusters As Telescopes (CATS) team \citep{Jauzac_2016}. The simply parameterized method assigns dark-matter halos to individual cluster galaxies and to the cluster, and uses halos with simple, physically motivated profiles each described by a small number of parameters. These models use the positions of multiply-imaged galaxies, and assign the mass to each galaxy halo using a proxy such as the stellar mass. \citet{Williams_2019} instead make no assumptions about the connection between luminous and dark matter, and use only the positions of multiply-imaged galaxies as constraints. They apply a ``free-form" approach which involves a large number of components that are not associated with any cluster galaxies but instead only uses the strong lensing image positions as input. \citet{Bradac_2009} created a free-form model which uses both strong and weak lensing image positions. A hybrid model was created by \citet{Diego_2015}, which uses a free-form approach to model the overall cluster halo and a parametric approach for the individual cluster members. \citet{Zitrin_2013} constructed a ``light-traces-mass" (LTM) model, which reconstructs the mass distribution of the cluster by smoothing and rescaling the surface brightnesses of the individual cluster members. \citet{Zitrin_2021} also constructed a parametric model using a Navarro, Frenk, \& White \citep[NFW;][]{Navarro_1996} density profile. We designate the two Zitrin models as ``Zitrin-LTM" and ``Zitrin-NFW."

The strongly lensed host galaxy of SN Refsdal and Icarus presents an opportunity to constrain the cluster lens' magnification at positions near the critical curve of the MACS\,J1149 cluster, which has not been possible for lens models of clusters. Known as Sp1149, the host galaxy is a triply-imaged and highly magnified face-on spiral galaxy at  $z=1.49$ \citep{Smith_2009,Teodoro_2018}. The three images of this galaxy are some of the largest images ever observed of a spiral at $z>1$ (see Fig.~\ref{fig:macsj1149}). The high magnification and relative lack of image distortion make it possible to study the spatial structure of the galaxy in detail. In 2011, \cite{Yuan_2011} acquired integral field unit (IFU) spectroscopy of the largest image of Sp1149 with the OH-Suppressing Infra-Red Spectrograph \citep[OSIRIS;][]{Larkin_2006} on the Keck II 10\,m telescope. The H$\alpha$ map measured from these observations revealed more than ten resolved \ion{H}{2} regions located less than 10$\arcsec$ from the critical curve (see Fig.~\ref{fig:osiris_HAmap}).

The \ion{H}{2} regions in Sp1149 should, in principle, allow a direct measurement of the magnification due to the MACS\,J1149 cluster by utilizing the empirical relationship between Balmer luminosities and velocity dispersions of \ion{H}{2} galaxies and giant \ion{H}{2} regions \citep[the $L-\sigma$ relation;][]{Terlevich_Melnick_1988}. The $L-\sigma$ relation for \ion{H}{2} galaxies has been shown to be consistent with luminosity distances expected for standard cosmological parameters to $z \approx 4$, and observations of \ion{H}{2} galaxies have been used to measure H$_0$ and constrain the dark energy equation of state \citep[e.g.,][]{Chavez_2016,Fernandez_Arenas_2018,Tsiapi_2021}. \cite{Terlevich_2016} used the $L-\sigma$ relation to estimate the intrinsic H$\beta$ luminosity of a single compact \ion{H}{2} galaxy at $z=3.12$ that is gravitationally lensed by the HFF cluster Abell S0163, and inferred a magnification of $23 \pm 11$, in agreement with the value of $\sim 17$ predicted by a simply parameterized model presented by \citet{Caminha_2016}. Due to their compact size and intrinsic faintntess, there are few constraints on the $L-\sigma$ relation for giant \ion{H}{2} regions at redshifts beyond $z\approx1$. If the $L-\sigma$ relation measured at low redshift is an accurate description of the giant \ion{H}{2} regions in Sp1149 at $z\approx 1.5$, the Balmer luminosities of the \ion{H}{2} regions can be used as standardizable candles and the magnification due to lensing at their positions can be constrained. Previously, direct measurements of a galaxy cluster's magnification have only been made using SNe~Ia at offsets of more than 20$\arcsec$ from the critical curves where the magnifications are $\lesssim 2$ \citep{Nordin_2014,Patel_2014,Rodney_2015,Rubin_2018}.

In Paper I of this series, Williams et al. (2023, submitted), we used a combination of archival OSIRIS IFU data and newly acquired spectroscopy from the Mulit-Object Spectrometer For Infra-Red Exploration (MOSFIRE) to measure the H$\alpha$ luminosities and intrinsic velocity dispersions of 11 \ion{H}{2} regions in Sp1149. After correcting for magnification using the GLAFIC mass model (v3) of MACS\,J1149 \citep{Oguri_2010,Kawamata_2016}, we found that the \ion{H}{2} regions in Sp1149 were $6.4^{+2.9}_{-2.0}$ times more luminous than expected from the locally calibrated $L-\sigma$ relation. 
%If the GLAFIC model's magnification predictions are accurate, this result would suggest that the zero-point of the $L-\sigma$ relation has a more positive value in Sp1149. 
However, if we instead assume that the $L-\sigma$ relation for giant \ion{H}{2} regions calibrated using low-redshift galaxies accurately describes those in Sp1149, then this result would suggest that the GLAFIC model underpredicts the magnification at the positions of the \ion{H}{2} regions by a factor of $\sim6$.

In this work, we make the assumption that the $L-\sigma$ relation in Sp1149 is identical to that in low-redshift galaxies, and infer the magnification due to lensing at the positions of 11 \ion{H}{2} regions in Sp1149. %Since gravitational lensing amplifies flux but does not affect emission-line widths, the velocity dispersions of the \ion{H}{2} regions can be used to infer their intrinsic luminosities according to the $L-\sigma$ relation, and the magnification can be constrained.
The \ion{H}{2} regions are adjacent to the critical curve of the MACS\,J1149 cluster ($\sim 2''$--$8''$), where magnifications are expected to reach up to $\sim20$. We compare our magnification measurements with the predicted magnifications from ten different lens models of MACS\,J1149. Magnification depends on the second derivative of the gravitational potential, so these measurements test a different aspect of the cluster models than the relative time delays from SN Refsdal, which depend on the difference in the potential between the multiple appearances.

In Section~\ref{sec:obs} we describe the OSIRIS and MOSFIRE observations and data reduction. Section~\ref{sec:measurements} details our method for inferring the magnification due to lensing at the position of each \ion{H}{2} region. We compare our measurements to the magnifications predicted by the models in Section~\ref{sec:results} and discuss the implications of our results in Section~\ref{sec:conclusion}.

\section{Observations and Data Reduction}\label{sec:obs}

Rest-frame optical IFU spectroscopy of the largest image of Sp1149 was obtained by \cite{Yuan_2011} with the OSIRIS instrument on the Keck~II 10~m telescope \citep{Larkin_2006}. The observations were taken using the Hn3 filter (15,940--16,760~\AA; resolution $R \equiv \lambda\/\Delta\lambda \approx 3400$), capturing the H$\alpha$ emission line at the redshift of Sp1149. Adaptive optics (AO) provided a corrected spatial resolution of 0.1$\arcsec$, corresponding to $\sim 300$~pc in the source plane for a typical magnification of $\mu=8$. The total exposure time was 4.75~hr. Data-reduction details are described by \cite{Yuan_2011}. We extracted the one-dimensional (1D) spectra of 11 \ion{H}{2} regions in Sp1149, inside circular apertures with radius $r=500$~pc in the source plane, given the GLAFIC model magnification predictions. Figure~\ref{fig:osiris_HAmap} shows the H$\alpha$ emission-line intensity map of Sp1149 and the \ion{H}{2} region extraction apertures. 

To measure the Balmer decrement and infer the extinction due to dust at the positions of the \ion{H}{2} regions, we acquired multislit spectroscopy using MOSFIRE on the Keck~II 10~m telescope \citep{McLean} over two half nights, February 25 and 26, 2020 UTC. We observed each \ion{H}{2} region in the {\it J} and {\it H} bands to detect the H$\beta$ and H$\alpha$ emission lines, respectively, and total exposure times ranged from 8~min to 36~min on six different slit masks. 

The spectra were reduced with the MOSFIRE {\tt Data Reduction Pipeline} \citep[{\tt DRP};][]{Konidaris_2019}. We used observations of telluric standard stars to correct for telluric absorption and acquired spectra of field stars on each slit mask to measure the absolute flux calibration for each mask. See Paper I for a detailed description of the MOSFIRE observations and data reduction. 

\section{Magnification Measurements}\label{sec:measurements}

\begin{figure*}[ht]
    \centering
    \includegraphics[width=.85\linewidth]{ 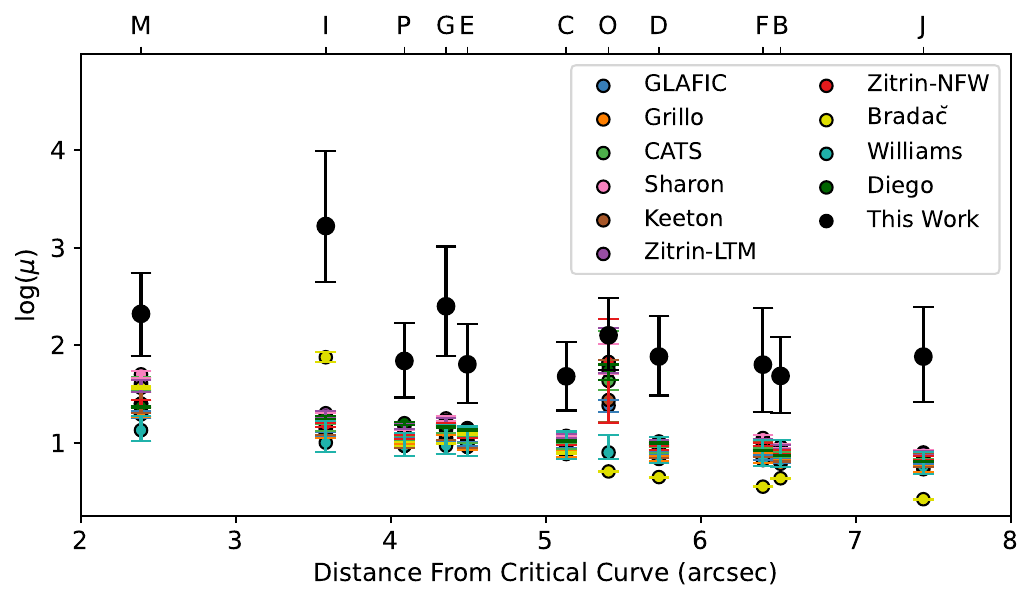}
    \caption{Magnification measurements for each of the observed \ion{H}{2} regions (large black points) overlaid on each of the models' predictions for the magnification at their positions (colored points).  Error bars correspond to 1$\sigma$ uncertainties.}
    \label{fig:magnifications}
\end{figure*}

\begin{table*}[hbt!]
\centering
\ra{1.6}
\caption{Magnification Measurements$^a$}
\scriptsize
\begin{tabular}{llllllllllll}

 & \small{Measured} & GLAFIC & Grillo & CATS & Sharon & Keeton & Zitrin$_{\tiny{\rm LTM}}$ & Zitrin$_{\tiny{\rm NFW}}$ &Brada\u{c} & Williams & Diego \\ 
 \midrule

\small{B} & \small{47.8$^{+77.4}_{-27.6}$} & 6.5$^{+0.3}_{-0.3}$ & 6.1$^{+0.5}_{-0.5}$ & 7.5$^{+0.6}_{-0.4}$ & 9.1$^{+0.5}_{-0.2}$ & 7.5$^{+1.2}_{-1.1}$ & 9.1$^{+0.4}_{-0.4}$ & 8.1$^{+0.3}_{-0.3}$ & 4.3$^{+0.0}_{-0.1}$ & 6.6$^{+4.0}_{-1.0}$ & 7.4$^{+0.1}_{-0.1}$\\
\small{C} & \small{47.9$^{+60.3}_{-26.8}$} & 8.4$^{+0.3}_{-0.4}$ & 7.6$^{+0.6}_{-0.5}$ & 9.1$^{+0.7}_{-0.5}$ & 11.2$^{+0.6}_{-0.4}$ & 8.9$^{+1.5}_{-1.2}$ & 11.8$^{+0.5}_{-0.6}$ & 9.9$^{+0.5}_{-0.4}$ & 8.0$^{+0.1}_{-0.3}$ & 8.5$^{+4.6}_{-1.7}$ & 10.5$^{+0.1}_{-0.3}$\\
\small{D} & \small{77.5$^{+124}_{-46.6}$} & 7.8$^{+0.3}_{-0.4}$ & 6.8$^{+0.5}_{-0.4}$ & 8.4$^{+0.6}_{-0.4}$ & 9.6$^{+0.5}_{-0.3}$ & 8.0$^{+1.3}_{-1.1}$ & 10.3$^{+0.3}_{-0.5}$ & 8.8$^{+0.4}_{-0.3}$ & 4.4$^{+0.1}_{-0.1}$ & 7.3$^{+4.0}_{-1.2}$ & 9.9$^{+0.1}_{-0.3}$\\
\small{E} & \small{63.1$^{+101}_{-37.3}$} & 9.7$^{+0.4}_{-0.5}$ & 9.0$^{+0.8}_{-0.7}$ & 10.6$^{+0.9}_{-0.6}$ & 13.2$^{+0.7}_{-0.5}$ & 10.3$^{+1.7}_{-1.5}$ & 14.1$^{+0.6}_{-0.9}$ & 11.7$^{+0.5}_{-0.5}$ & 12.3$^{+0.1}_{-0.3}$ & 9.3$^{+5.4}_{-2.0}$ & 13.5$^{+0.2}_{-0.4}$\\
\small{F} & \small{64.8$^{+167}_{-44.1}$} & 7.2$^{+0.3}_{-0.5}$ & 6.9$^{+0.9}_{-0.8}$ & 9.3$^{+1.0}_{-0.5}$ & 11.2$^{+0.8}_{-0.4}$ & 8.7$^{+1.7}_{-1.5}$ & 10.3$^{+0.6}_{-0.5}$ & 9.7$^{+0.3}_{-0.4}$ & 3.5$^{+0.1}_{-0.0}$ & 6.6$^{+4.0}_{-0.8}$ & 8.3$^{+0.1}_{-0.1}$\\
\small{G} & \small{252$^{+762}_{-171}$} & 11.3$^{+0.7}_{-0.7}$ & 10.9$^{+1.2}_{-1.1}$ & 13.6$^{+1.7}_{-0.9}$ & 17.8$^{+1.2}_{-0.9}$ & 12.4$^{+2.3}_{-1.8}$ & 16.8$^{+1.1}_{-1.5}$ & 15.3$^{+0.5}_{-0.8}$ & 9.7$^{+0.1}_{-0.1}$ & 9.3$^{+5.9}_{-1.6}$ & 14.6$^{+0.3}_{-0.4}$\\
\small{I} & 1.65$^{+7.97}_{-1.21}\tiny{\times 10^3}$ & 12.3$^{+0.7}_{-0.7}$ & 12.1$^{+1.1}_{-1.0}$ & 14.1$^{+1.5}_{-0.9}$ & 19.1$^{+1.2}_{-0.9}$ & 13.8$^{+2.7}_{-2.0}$ & 20.1$^{+1.0}_{-1.9}$ & 15.3$^{+0.8}_{-0.7}$ & 75.3$^{+10.6}_{-7.6}$ & 10.0$^{+6.5}_{-2.0}$ & 18.4$^{+0.4}_{-0.7}$\\
\small{J} & \small{76.5$^{+168}_{-50.2}$} & 5.9$^{+0.3}_{-0.3}$ & 5.3$^{+0.4}_{-0.4}$ & 7.2$^{+0.6}_{-0.4}$ & 7.9$^{+0.4}_{-0.2}$ & 6.6$^{+1.1}_{-1.0}$ & 7.8$^{+0.4}_{-0.3}$ & 6.9$^{+0.3}_{-0.3}$ & 2.6$^{+0.0}_{-0.0}$ & 5.7$^{+2.3}_{-0.9}$ & 6.4$^{+0.1}_{-0.1}$\\
\small{M} & \small{203$^{+343}_{-125}$} & 19.5$^{+1.7}_{-1.7}$ & 20.8$^{+2.7}_{-2.3}$ & 38.9$^{+8.0}_{-4.4}$ & 50.3$^{+4.7}_{-4.2}$ & 25.0$^{+8.0}_{-4.7}$ & 40.9$^{+3.3}_{-7.6}$ & 25.2$^{+2.0}_{-2.0}$ & 36.6$^{+1.2}_{-1.1}$ & 13.4$^{+5.3}_{-3.0}$ & 23.5$^{+0.3}_{-0.6}$\\
\small{O} & \small{129$^{+179}_{-72.6}$} & 24.1$^{+3.1}_{-3.3}$ & 39.4$^{+32.5}_{-24.5}$ & 43.0$^{+96.7}_{-8.0}$ & 65.3$^{+36.8}_{-15.1}$ & 27.4$^{+43.8}_{-11.0}$ & 63.5$^{+86.5}_{-12.1}$ & 67.3$^{+119.4}_{-51.5}$ & 5.1$^{+0.0}_{-0.0}$ & 7.9$^{+4.1}_{-1.2}$ & 55.3$^{+8.0}_{-11.5}$\\
\small{P} & \small{69.2$^{+97.2}_{-39.9}$} & 10.8$^{+0.4}_{-0.7}$ & 9.3$^{+0.5}_{-0.5}$ & 10.9$^{+0.8}_{-0.6}$ & 12.8$^{+0.7}_{-0.4}$ & 10.4$^{+1.9}_{-1.4}$ & 14.8$^{+0.9}_{-0.9}$ & 11.3$^{+0.7}_{-0.5}$ & 9.8$^{+0.4}_{-0.5}$ & 9.3$^{+3.2}_{-1.9}$ & 15.8$^{+0.4}_{-0.7}$\\
\midrule
\multicolumn{2}{c}{\small{$\langle \mu_{\rm obs}$/$\mu_{\rm mod} \rangle$}} & 7.0 $\pm$ 2.4 & 7.1 $\pm$ 2.4 & 6.3 $\pm$ 2.2 & 4.1 $\pm$ 1.4 & 6.9 $\pm$ 2.4 & 4.8 $\pm$ 1.7 & 5.8 $\pm$ 2.0 & 7.4 $\pm$ 2.7 & 8.3 $\pm$ 2.9 & 4.1 $\pm$ 1.6\\
\multicolumn{2}{c}{Median Tension}&2.3~$\sigma$ & 2.3~$\sigma$ & 2.1~$\sigma$ & 1.9~$\sigma$ & 2.1~$\sigma$ & 1.8~$\sigma$ & 2.1~$\sigma$ & 2.6~$\sigma$ & 2.3~$\sigma$ & 2.1~$\sigma$  \\
\bottomrule

\end{tabular}
{\footnotesize $^a$Model-predicted magnifications at the position of each \ion{H}{2} region, and our measured magnifications. The ratio $\mu_{\rm obs}$/$\mu_{\rm mod}$ is the weighted average of the ratios of the measured to model-predicted magnifications for the 11 \ion{H}{2} regions. Reported uncertainties are 1$\sigma$. These measurements are shown in Figure~\ref{fig:magnifications}.}
\label{tab:magnifications}
\end{table*}

To infer magnification values using the Balmer $L-\sigma$ relation, we require a calibration of the $L-\sigma$ relation that uses aperture sizes and spectral resolution comparable to those of our OSIRIS measurements of the H$\alpha$ luminosities and velocity dispersions of the \ion{H}{2} regions in Sp1149. Using archival IFU spectroscopy of nine nearby spiral galaxies taken with the Multi Unit Spectroscopic Explorer \citep[MUSE;][]{Bacon_2010} on the Very Large Telescope (VLT), we extract the spectra of 347 \ion{H}{2} regions at $z\approx 0$ using the same physical aperture sizes, given the GLAFIC model predictions, that we used to extract the \ion{H}{2} regions in Sp1149 from the OSIRIS data. 

 \begin{figure*}[ht]
    \centering
    \includegraphics[width=.85\linewidth]{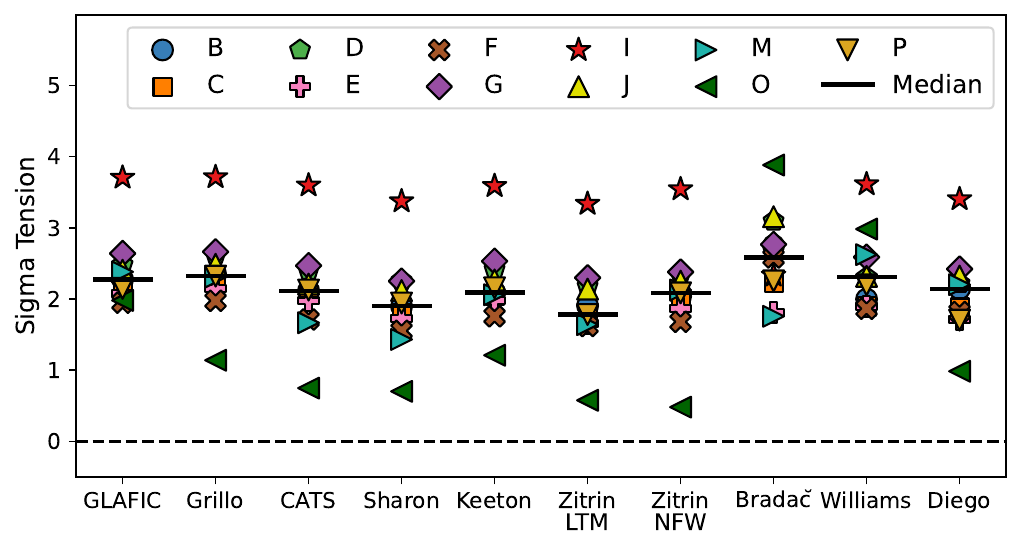}
    \caption{The statistical tension between our measurements of the magnification compared with each model’s prediction for the magnification at the position of each \ion{H}{2} region. A positive value for the tension indicates that our measured magnification is greater than the model's prediction. The black lines indicate the median tension for each model. We find that all ten models underpredict the magnification compared to the measurements by a median value among the \ion{H}{2} regions of 1.8--2.5$\sigma$.}
    \label{fig:sigma_tension}
\end{figure*}

We employ Markov-Chain Monte Carlo (MCMC) sampling with the {\tt pymc3} package \citep{pymc3} to measure the intrinsic velocity dispersion and H$\alpha$ luminosity of each \ion{H}{2} region in the local sample from the H$\alpha$ and H$\beta$ emission lines. From the posteriors for all of the \ion{H}{2} regions, we apply a hierarchical linear mixture model implemented by the {\tt linmix} software \citep{Kelly_2007} to infer the slope and intercept of the $L-\sigma$ relation at low redshift,

\begin{equation}\label{equation:Lsigma}
    \log(L_{\rm{H}\alpha}) = (0.85\pm0.11) \log(\sigma) + (38.46\pm0.13)\, ,
\end{equation}
\noindent where $L_{{\rm H}\alpha}$ is in units of erg~s$^{-1}$ and $\sigma$ is in units of km~s$^{-1}$. The root-mean-square intrinsic scatter is 0.34~dex. Additional descriptions of our  calibration of the relation at low redshift may be found in Paper I. 

In Paper I, we presented measurements of the intrinsic H$\alpha$ luminosity and velocity dispersion of 11 \ion{H}{2} regions in Sp1149. We used our MOSFIRE observations of the H$\beta$ and H$\alpha$ emission lines for each \ion{H}{2} region to constrain the extinction due to dust, and the OSIRIS observations of the H$\alpha$ emission lines to infer their intrinsic velocity dispersions and, in combination with our constraints on extinction, intrinsic H$\alpha$ luminosities. 

To measure the magnification at the position of each \ion{H}{2} region, we use the posteriors from Paper I to compute the observed (magnified) H$\alpha$ luminosity, $L_{\rm obs}({\rm H\alpha})$. We compute the expected H$\alpha$ luminosity, $L_{\rm exp}({\rm H\alpha})$ of each \ion{H}{2} region using the posteriors for the velocity dispersion from Paper I and applying our calibration for the local $L-\sigma$ relation (Eq.~\ref{equation:Lsigma}). The posteriors on the slope and intercept of the low-redshift relation are used to propagate the uncertainties associated with the calibration. The observed magnification, $\mu_{\rm obs}$, at the position of each \ion{H}{2} region is given by 

\begin{equation}
    \log(\mu_{\rm obs}) = L_{\rm obs}({\rm H\alpha}) - L_{\rm exp}({\rm H\alpha})\, .
\end{equation}
\noindent
We list our measurements of the magnifications at the positions of 11 \ion{H}{2} regions in Sp1149 in Table~\ref{tab:magnifications}, and show these as a function of distance from the critical curve of MACS\,J1149 at $z=1.49$ in Figure~\ref{fig:magnifications}.

\section{Comparison with the Models}\label{sec:results}

\begin{figure*}[ht]
    \centering
    \includegraphics[width=.9\linewidth]{ 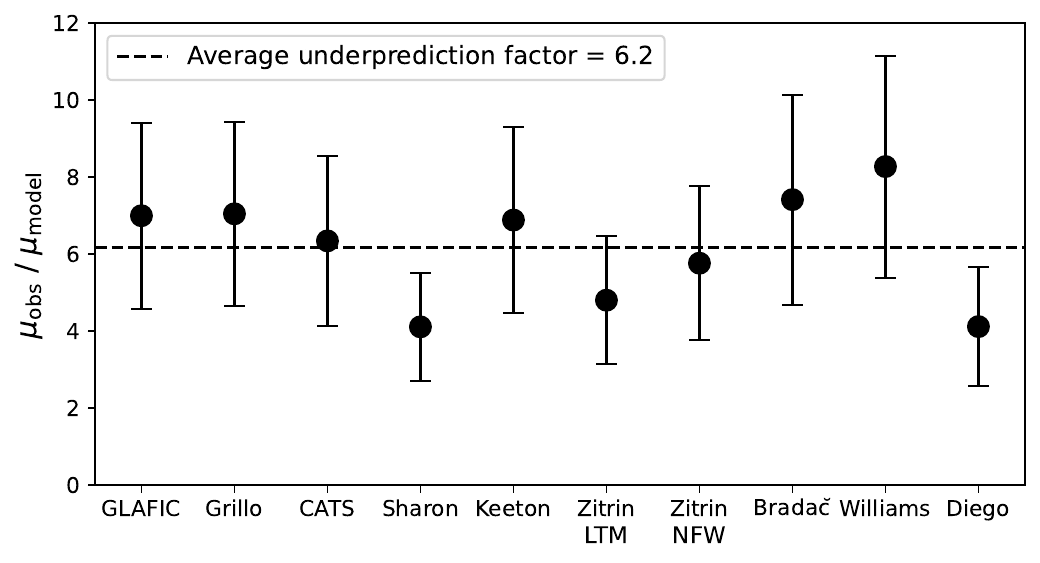}
    \caption{The factor by which each model underpredicts the magnification of the \ion{H}{2} regions in Sp1149. The average underprediction factor for all the models is 6.2).}
    \label{fig:factors}
\end{figure*}

We next compare our observed magnifications of the 11 \ion{H}{2} regions in Sp1149 with the magnifications predicted by ten different models of the MACS\,J1149 cluster. Each lens modeling team provided $\sim 100$ model magnification maps corresponding to different MCMC realizations\footnote{https://archive.stsci.edu/pub/hlsp/frontier/macs1149/models/}. We use the median value at the position of each \ion{H}{2} region as the predicted magnification, adopting the 16th and 84th percentile values to compute the 1$\sigma$ uncertainties. The model magnification values are listed in Table~\ref{tab:magnifications}, and the model predictions together with our constraints are plotted in Figure~\ref{fig:magnifications}.

To evaluate the level of agreement between our measurements and the model predictions, we calculate the tension between each measurement and prediction. We take the difference between the measurement and the prediction, and divide by the $1\sigma$ uncertainty in the difference. A positive tension indicates that the value we measure is greater than the model's prediction. 

In Figure \ref{fig:sigma_tension}, we plot the tension for each model and \ion{H}{2} region, as well as the median tension for each model. 
Almost all of our magnification measurements are $\sim 1$--3$\sigma$ greater than the models' predictions. The median statistical tension among the set of 11 \ion{H}{2} regions for each model is 1.8--2.6$\sigma$.  

We next compute the average factor by which each model underpredicts the magnification compared to our measurements by computing $\mu_{\rm obs} / \mu_{\rm model}$ for each \ion{H}{2} region and assigning a weight to each measurement based on the combined 1$\sigma$ uncertainties of the observed magnification and the model prediction. As Figure \ref{fig:factors} shows, the 10 available models of MACS\,J1149 underpredict the magnification by a factor of $\sim 5$--8. The average underprediction factor among the models is $6.2$, and the median factor among the models is $6.6$. 

\begin{figure}
    \centering
    \includegraphics[width=\linewidth]{ 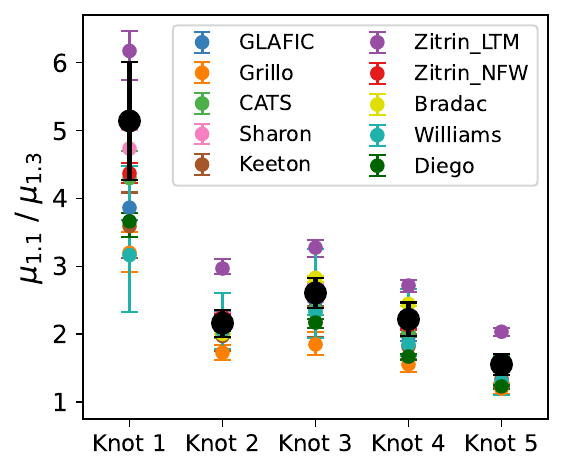}
    \caption{The flux ratios of five bright knots in Image 1.1 and Image 1.3 of Sp1149, measured from the F606W {\it HST} imaging of the MACS \,J1149 cluster field. We compare these flux ratios to the predicted magnification ratios at their positions and find that the measured ratios agree with the predictions from the models within the 1$\sigma$ uncertainties. }
    \label{fig:knot_ratios}
\end{figure}

\begin{figure*}
    \centering
    \includegraphics[width=.8\linewidth]{ 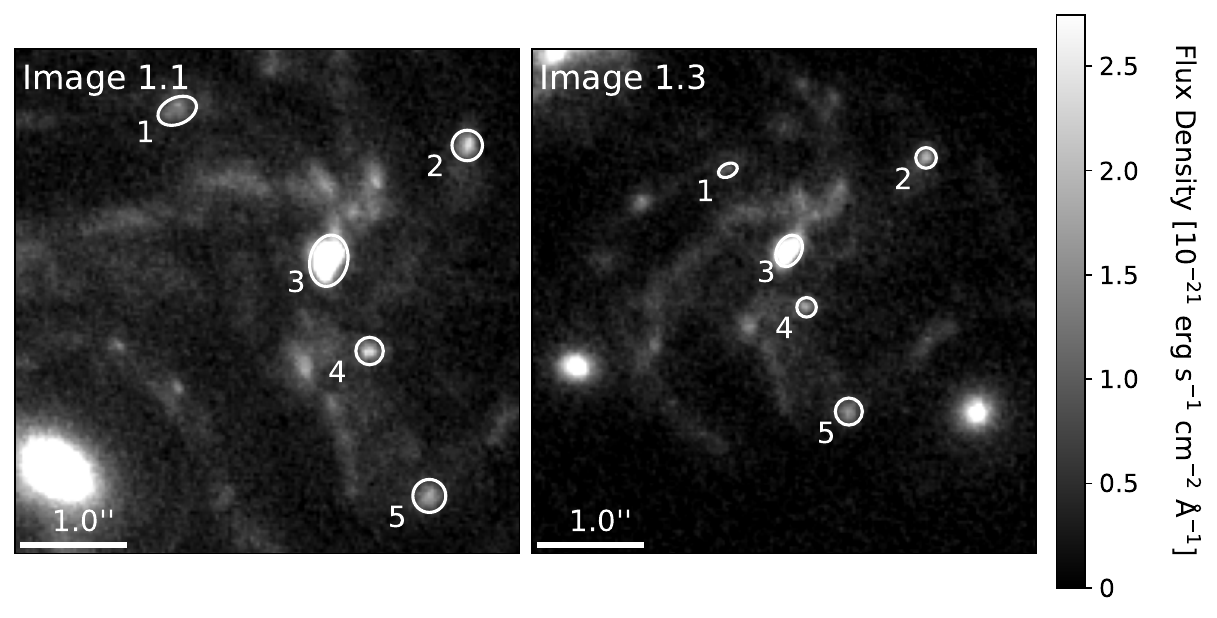}
    \caption{{\it HST} F606W close-up images of Image 1.1 and Image 1.3 of Sp1149, with five bright knots identified in each image. We measure the flux density of each knot in both images and compare their flux ratios to the magnification ratios predicted by each model. The white regions show the apertures used to measure the flux of each knot.}
    \label{fig:knots}
\end{figure*}

\section{Discussion}\label{sec:conclusion}
Using the Balmer $L-\sigma$ relation, we measured the magnification due to gravitational lensing by the MACS\,J1149 cluster of 11 giant \ion{H}{2} regions in the spiral galaxy Sp1149 ($z=1.49$). We have compared our measurements to the magnifications predicted by 10 different cluster mass models and found that all of the models predict magnifications that are smaller than our inferred values at the positions of the \ion{H}{2} regions by factors of $\sim 5$--8. The tension between our measurements and the model predictions is 1.8--2.6$\sigma$.

The models that are in the least tension with our measurements ($\sim1.8\sigma$) are the simply parameterized Sharon model and the Zitrin light-traces-mass model, which underpredict the magnification in comparison to our constraints by an average factor of $4.1\pm1.4$ and $4.8\pm1.7$, respectively. The Williams free-form model underpredicts the magnification by the largest average factor among the models, with $\langle \mu_{\rm obs}/\mu_{\rm model}\rangle=8.3\pm2.9$. 

Here we have calculated the magnification under the assumption that the $L-\sigma$ relation for giant \ion{H}{2} regions that we have calibrated at low redshift in matching apertures and spectral resolution applies to the \ion{H}{2} regions in SN Refsdal's host galaxy at $z=1.49$. Previous studies have shown that the $L-\sigma$ relation for \ion{H}{2} galaxies does not evolve strongly with redshift at least $z\approx 4$, and it has been used to calculate H$_0$ and the dark energy equation of state \citep[e.g.,][]{Chavez_2016,Fernandez_Arenas_2018,Tsiapi_2021}. If we instead assume that the magnifications predicted by the models are accurate, then our results indicate that \ion{H}{2} regions are substantially more luminous in Sp1149 than predicted by the low-redshift $L-\sigma$ realtion (3$\sigma$ tension), and would imply a physical difference between the two populations of \ion{H}{2} regions (see Paper I).  
%Similar measurements of magnified \ion{H}{2} regions at $z\gtrsim1$ that are lensed by different clusters would be necessary to determine whether our results are indicative of a redshift evolution in the $L-\sigma$ relation or a systematic bias in the MACS\,J1149 cluster models. 

If our results correspond to a systematic bias of the lens models of the MACS\,J1149 cluster, the magnification corrections applied to background galaxies that are lensed by MACS\,J1149 would cause us to overestimate their intrinsic luminosities and emission-line fluxes. Physical properties of these galaxies, such as star-formation rate and stellar mass, would also be overestimated. The value of H$_0$ inferred from time-delay measurements is sensitive to the details of cluster models, so a systematic bias in the lens models used to infer H$_0$ from SN Refsdal could affect the interpretation of that measurement. For instance, increasing the magnification by a factor of 6 would require a mass sheet of $\kappa=0.6$ under the mass-sheet degeneracy \citep[$\mu \propto (1-\kappa)^{-2}$;][]{Oguri_2003}, which would reduce the derived value of H$_0$ by a factor of $1-\kappa=0.4$. The value of H$_0$ inferred from the time delays of SN Refsdal is $64.8^{+4.4}_{-4.3}$ km s$^{-1}$ Mpc$^{-1}$, so this interpretation would imply an implausible H$_0\approx26$~km~s$^{-1}$~Mpc$^{-1}$.

To test the likelihood of the magnification of Image 1.1 of Sp1149 being a factor of $\sim3$ times higher than the models predict, we compare the observed flux ratios of bright knots in Image 1.1 and Image 1.3 to the model-predicted magnification ratios at their positions. As shown in Figure~\ref{fig:macsj1149}, Image 1.3 is substantially farther from the critical curve than mirrored Images 1.1 and 1.2. %where magnifications are smaller and models tend to be more accurate. 
We identify five bright knots in Sp1149 and measure their flux densities in both Image 1.1 and Image 1.3 from the {\it HST} F606W image of MACS\,J1149 (see Figure~\ref{fig:knots}). We find that, for all five knots, the observed flux ratios agree with the predicted magnification ratios from a majority of the models (see Figure~\ref{fig:knot_ratios}).
In other galaxy-cluster fields, the magnifications of SNe~Ia have been measured at similar offsets from the critical curve of $\sim 20''$, and obtained approximate agreement \citep{Nordin_2014,Patel_2014,Rodney_2015,Rubin_2018}. Additionally, a factor of 6 bias in the predicted magnifications in Image 1.1 would imply an improbably small value of H$_0\approx26$~km~s$^{-1}$~Mpc$^{-1}$.
Consequently, we conclude that the discrepancy we identify between the predicted and measured H$\alpha$ luminosities of the giant \ion{H}{2} regions in Sp1149 is most likely not due to a systematic problem in the lens models.  Instead, the Balmer $L-\sigma$ relation of the \ion{H}{2} regions in Sp1149 is likely offset to higher luminosities by a factor of $\sim 6$. Observations of the luminosities and velocity dispersions of \ion{H}{2} regions in magnified galaxies are needed to confirm this conclusion. 

The Balmer $L-\sigma$ relation is a well-established empirical correlation, but the physical origin of this relation is not yet understood. While the $L-\sigma$ relation is constant with redshift for \ion{H}{2} galaxies out to at least $z\approx 4$, our results suggest that it may not apply for individual \ion{H}{2} regions at $z\gtrsim1$.

%If we assume that the magnification predictions for image 1.3 are accurate, then the observed flux ratios of the knots in image 1.1 compared to those in image 1.3 would be a factor of $\sim3$ higher than the model-predicted magnification ratios if our $L-\sigma$ measurements represent a systematic under-prediction of the magnification in image 1.1.

\section{Acknowledgments}\label{sec:acknowledgments}

The lens models used in this work were downloaded from the HST Frontier Fields Data Access Page. 
Some of the data presented herein were obtained at the W. M. Keck
Observatory, which is operated as a scientific partnership among the
California Institute of Technology, the University of California, and
the National Aeronautics and Space Administration
(NASA); the observatory was made possible by the generous financial
support of the W. M. Keck Foundation.
We thank Josh Walawender and Sherry Yeh for their support with the MOSFIRE observations and data reduction, and Tiantian Yuan for providing the reduced OSIRIS datacube. We obtained Keck data from telescope time allocated to A.V.F. through the University of California, and  to P.L.K. through NASA [Keck PI Data Award \#1644110 (75\/2020A\_N110), administered by the NASA Exoplanet Science Institute through the agency's scientific partnership with the California Institute of Technology and the University of California].  We recognize the Maunakea summit as a sacred site within the indigenous Hawaiian community and we are grateful for the opportunity to conduct observations there. 

P.L.K. acknowledges funding from NSF grants AST-1908823 and AST-2308051. A.V.F. was supported by the Christopher R. Redlich Fund and many individual donors.

\bibliography{main}{}
\bibliographystyle{aasjournal}

\end{document}